\documentclass[journal]{juan}
\usepackage[T1]{fontenc}
\usepackage[latin1]{inputenc}
\usepackage{longtable} 
\usepackage{supertabular}
\usepackage{graphicx}
\usepackage{amssymb}
\usepackage{amsmath}
\usepackage{units}
\usepackage{natbib}
\usepackage{version}
\usepackage{color}
\usepackage{float}

\def\be{\begin{equation}}
\def\ee{\end{equation}}
\def\bi{\begin{itemize}}
\def\ei{\end{itemize}}
\newcommand{\Iris}{{\sc Iris}}
\newcommand{\Iras}{{\sc Iras}}
\newcommand{\Archeops}{{\sc Archeops}}
\newcommand{\Wmap}{{\sc Wmap}}

\newcommand{\healpix}{{\sc HEALpix}}

\begin{document}

\title{A characterization of the diffuse Galactic emissions in the anti-center of the Galaxy} 

\author{ L.~Fauvet$^{1,2}$, J.~F.~Mac\'{\i}as-P\'erez$^{2}$, S. R. Hildebrandt$^{2,3}$, F.-X. D\'esert$^{4}$}
\maketitle
\thanks{(1) European Space Agency (ESA), Research and Scientific Support Dpt., Astrophysics Division, Keplerlaan 1, 2201AZ Noordwijk, The Netherlands}\\
\thanks{(2) LPSC, Universit\'e Joseph Fourier Grenoble 1, CNRS/IN2P3, Institut National Polytechnique de Grenoble, 53 avenue des Martyrs, 38026 Grenoble cedex, France}\\
\thanks{(3) California Institute of Technology, 1200 E. California Bvd., 91125 Pasadena, USA}\\
\thanks{(4) Laboratoire d'astrophysique de Grenoble, IPAG, Universit\'e Joseph Fourier BP 53, 38041 GrenobleCEDEX 9, France} \\

\abstract Using the Archeops and WMAP data we perform a study of the anti-center Galactic diffuse emissions -- thermal dust,
synchrotron, free-free and anomalous emission -- at degree scales. The high frequency data are used to infer the thermal dust electromagnetic
spectrum and spatial distribution allowing us to precisely subtract this component at lower frequencies.
After subtraction of the thermal dust component a mixture of standard synchrotron and free-free emissions
does not account for the residuals at these low frequencies. Including the all-sky 408~MHz Haslam data we find evidences for  anomalous emission with a 
spectral index of -2.5 in $T_{RJ}$ units. However, we are
not able to conclude regarding the nature of this  anomalous emission in this region. For the purpose, data between 408 MHz and 20 GHz covering the same sky
region are needed.

\keywords ISM: general -- ISM: clouds -- Methods: data analysis -- Cosmology: observations -- Submillimeter -- Catalogs 

\date{\today}







\section{Introduction}
\begin{table*}
\begin{center}
\caption{rms of the high frequency data and of the residuals after subtraction of the dust model.  \label{tab:dustmodelresiduals}}
\vspace{0.3cm}

\begin{tabular}{|c|c|c|c|} \hline
 Frequency (GHz) &  Data rms  (mK$_{RJ}$) & Residual rms (mK$_{RJ}$) &  Noise standard deviation   (mK$_{RJ}$)  \\
 \hline
143   &   0.194407  &  0.0240274  &  0.0241575 \\
 217   &  0.315175  &  0.0306455   & 0.0406982 \\
 353   &  0.498923   & 0.0705655   & 0.0404479 \\
 545   &  0.699145   &  0.113738    & 0.157994 \\ 
3000  &  0.0973817  & 0.00830546 &   0.0197808 \\
5000 &  0.0134645 &  0.00974734  & 0.00176314 \\
\hline
\end{tabular}
\end{center}
\end{table*}
\indent The anomalous microwave emission (AME in the following), is an
important contributor of the Galactic diffuse emissions in the range from
20 to 60 GHz. It was first observed by \citep{1997ApJ...482L..17D,1996ApJ...464L...5K} and then identified by~\citep{leitch} as
 free-free emission from electrons with temperature, $T_e > 10^6$K. \cite{draine1998}
argued that AME may result from electric dipole radiation due to small
rotating grains, the so-called \emph{spinning dust}. Models of the \emph{spinning dust} emission \citep{draine1998b} 
show an electromagnetic spectra peaking at around 20-50 GHz being
able to reproduce the observations~\citep{finkbeiner2003,costa2004, watson, iglesias2005, cassasus, casassus2008,dickinson2009, tibbs}. The initial \emph{spinning dust}  model has been
refined regarding the shape and rotational properties of the dust
grains \citep{ali, hoang2010, hoang2011, silsbee}. An alternative explanations of AME was proposed by
\cite{draine1999} based on magnetic dipole radiation arising from hot
ferromagnetic grains. This kind of models associated to single-domain
predict polarization fraction much bigger than the electric dipole ones
~\citep{lazarian}. Original models have been mainly ruled out by many studies
\citep{batistelli, cassasus, kogut, mason, lopez} although modern variants of those
 may still be of interest (B. Draine private communication).\\

\noindent Correlation between microwave and infrared maps, mainly dominated by dust thermal emission \citep{desert90},
was observed for various experiments, for example
on COBE/DMR~\citep{kogut96a, kogut96b}, OVRO~\citep{leitch},
Saskatoon~\citep{costa1997}, survey at 19GHz~\citep{costa1998},
Tenerife~\citep{costa1999}. Similar signal was find in small region
by~\citep{finkbeiner2003} and in some molecular clouds based on data from
COSMOSOMAS~\citep{genova, watson}, AMI~\citep{scaife1, scaife2},
CBI~\citep{cassasus, castellanos}, VSA~\citep{tibbs} and
Planck~\citep{early}. Recent studies based on several sets of data
\citep{bot} found similar results.\\

\indent Independently, \cite{bennett} proposed an alternative explanation of AME based on
flat-spectrum synchrotron emission associated to star-forming
regions to explain part of the WMAP first-year observations. This
hypothesis seems to be disagreement with results from~\cite{costa2004,fernandez, hildebrandt2007, ysard} which showed that spinning dust was the
most trustable emission to explain the excess below 20 GHz. Furthermore,
\cite{davies} showed the existence of important correlation between microwave and
infrared emission in regions outside star-forming areas. More recently,
\cite{kogut2011} discussed the fact that \emph{spinning dust} fits better to
ARCADE data (3.8 and 10 GHz)  than a flat-spectrum synchrotron.\\

\indent We propose here to study the Galactic diffuse emissions in the
Galactic plane, particularly focusing on the anti-center region. 
The observational data, from 408~MHz to 3000 GHz, used for this study are presented in Section~\ref{data}.
Section \ref{dust} discusses in details the contribution of the 
diffuse Galactic thermal dust emission using the high frequency data. In Section~\ref{4_ff_model} we
consider a simple free-free and canonical synchrotron emission model for the
thermal dust subtracted microwave data. The possible contribution from anomalous emission is
discussed in Section~\ref{sync}. We draw conclusions in Section~\ref{conc}.




\begin{figure*}
\hspace*{-1.5cm}
\centering

\includegraphics[angle=90,height=5cm,width=7cm]{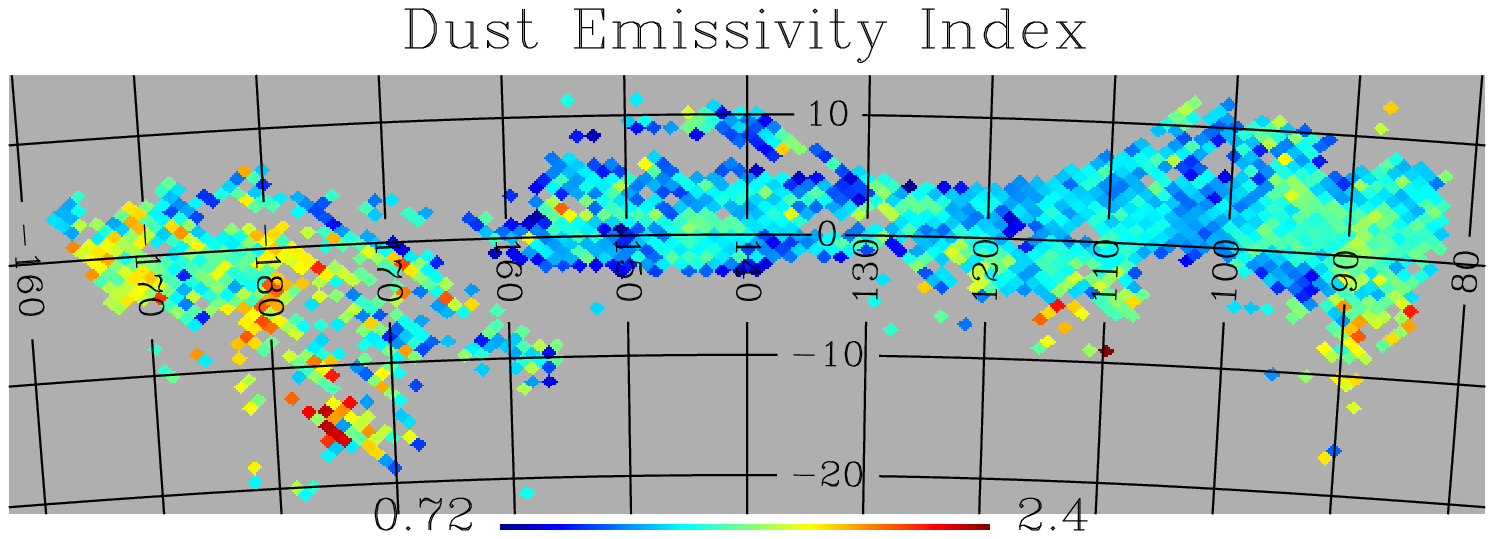}
\includegraphics[angle=90,height=5cm,width=7cm]{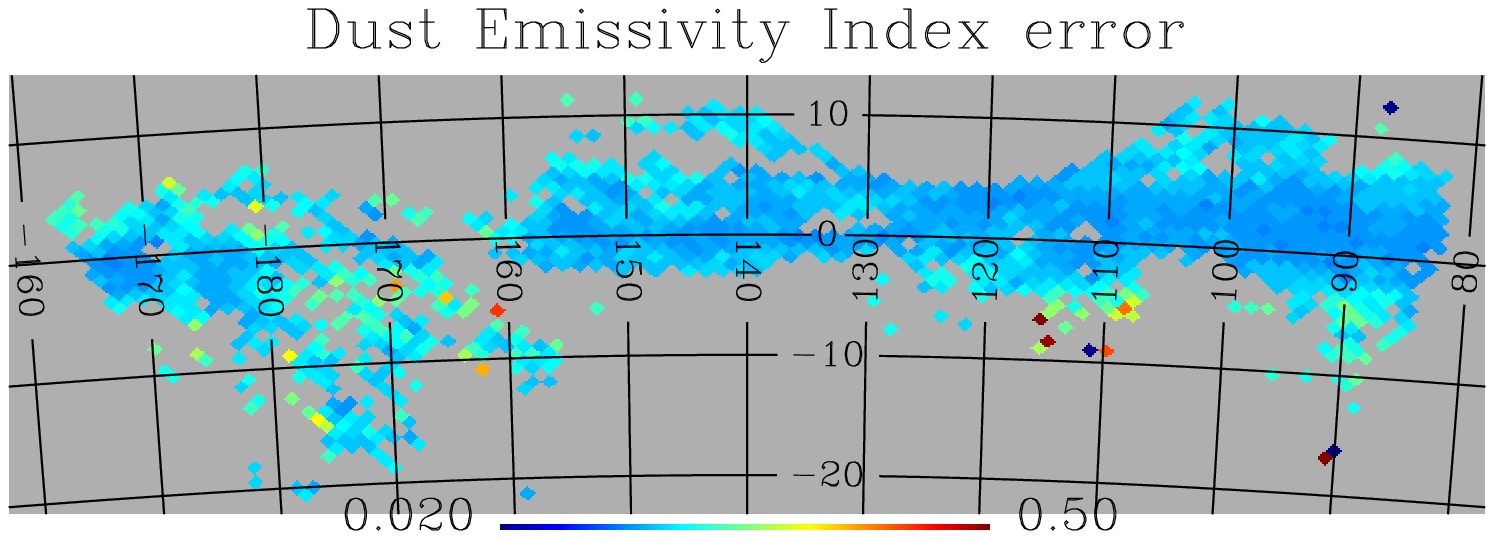}
\includegraphics[angle=90,height=5cm,width=7cm]{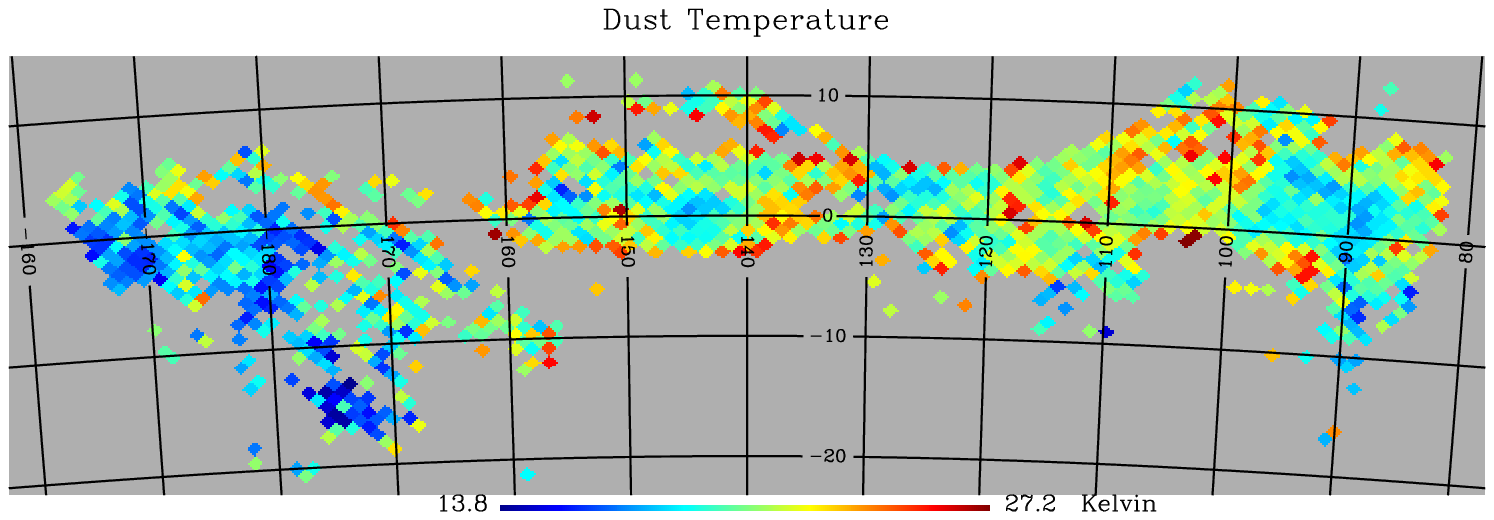}
\includegraphics[angle=90,height=5cm,width=7cm]{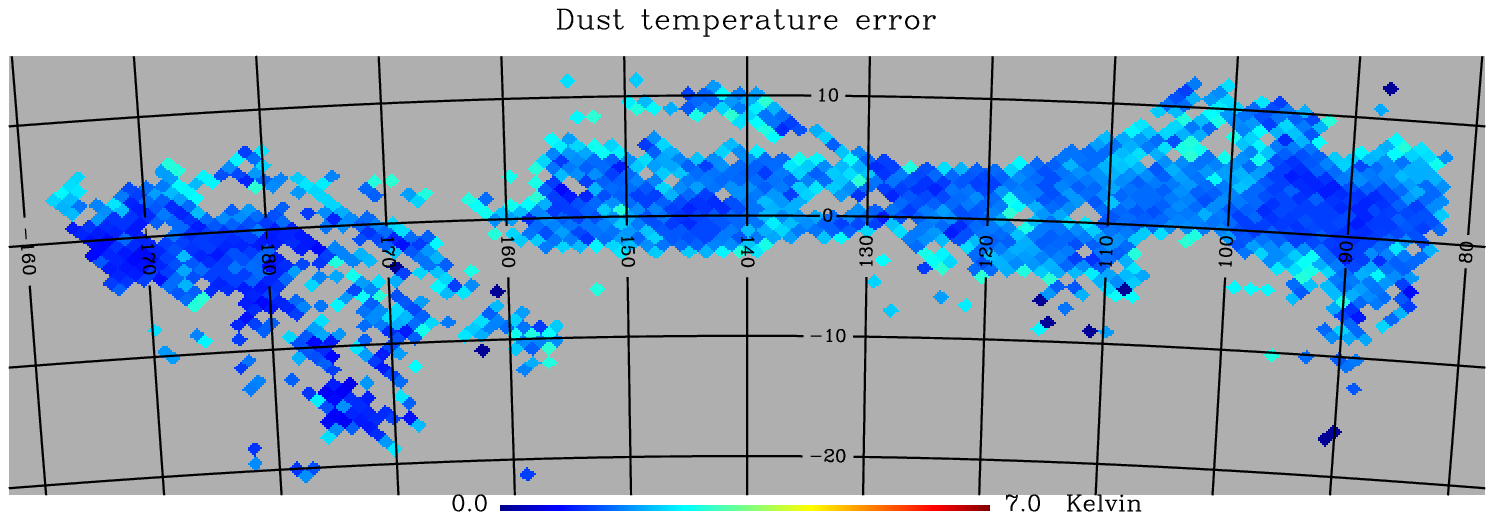}

\caption{From top to bottom and from left to right: maps of the best-fit thermal dust emission spectral indices and temperature  and uncertainties at 2 $\sigma$ (95 \% C.L.) \emph{(right)}. \label{fig:chp4_Tdust_map}}
\end{figure*}

\section{Microwave and millimeter observations}
\label{data}

\begin{figure*}[!th]
\centering
\includegraphics[angle=90,height=4cm,width=6cm]{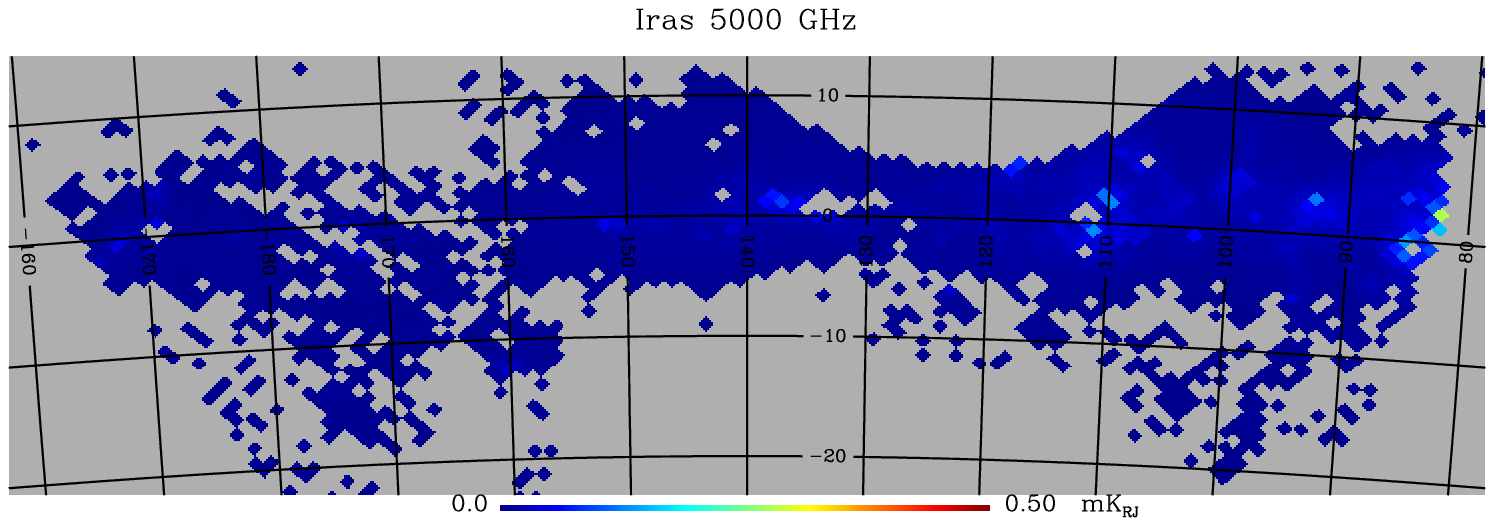}\includegraphics[angle=90,height=4cm,width=6cm]{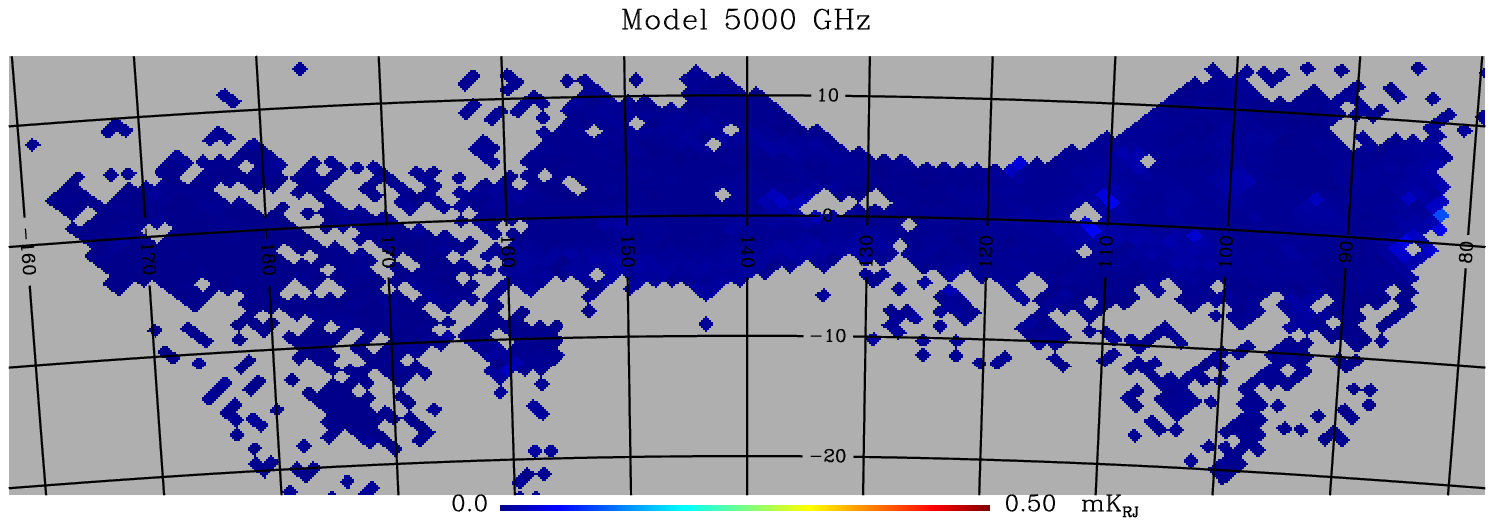}\includegraphics[angle=90,height=4cm,width=6cm]{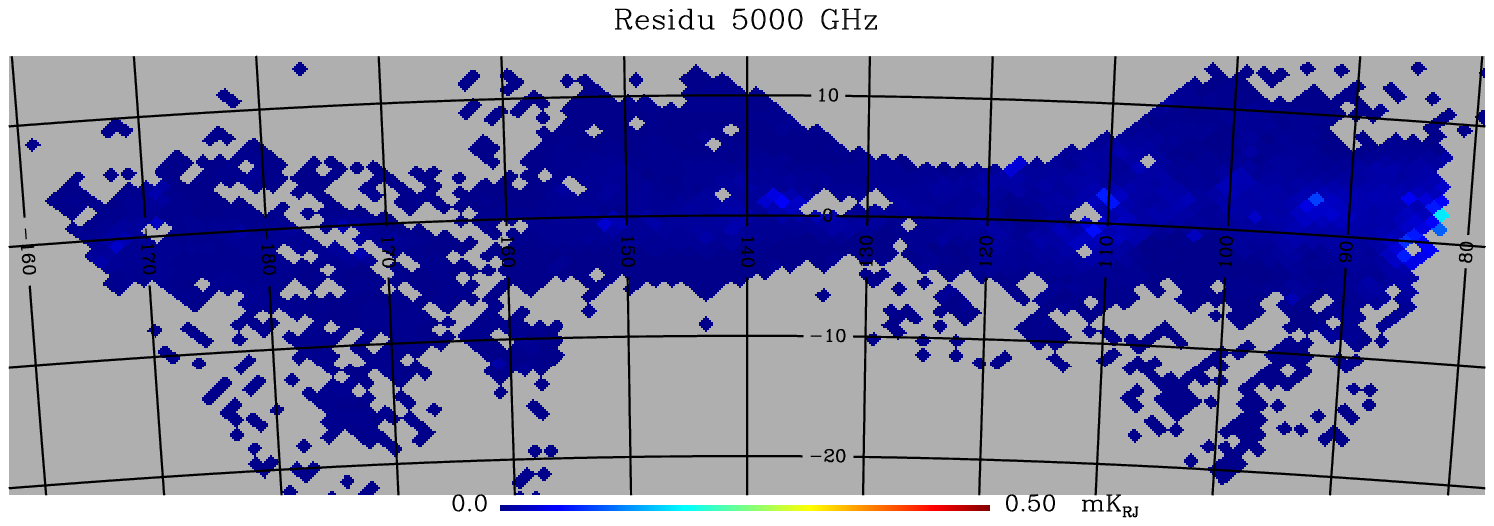}
\includegraphics[angle=90,height=4cm,width=6cm]{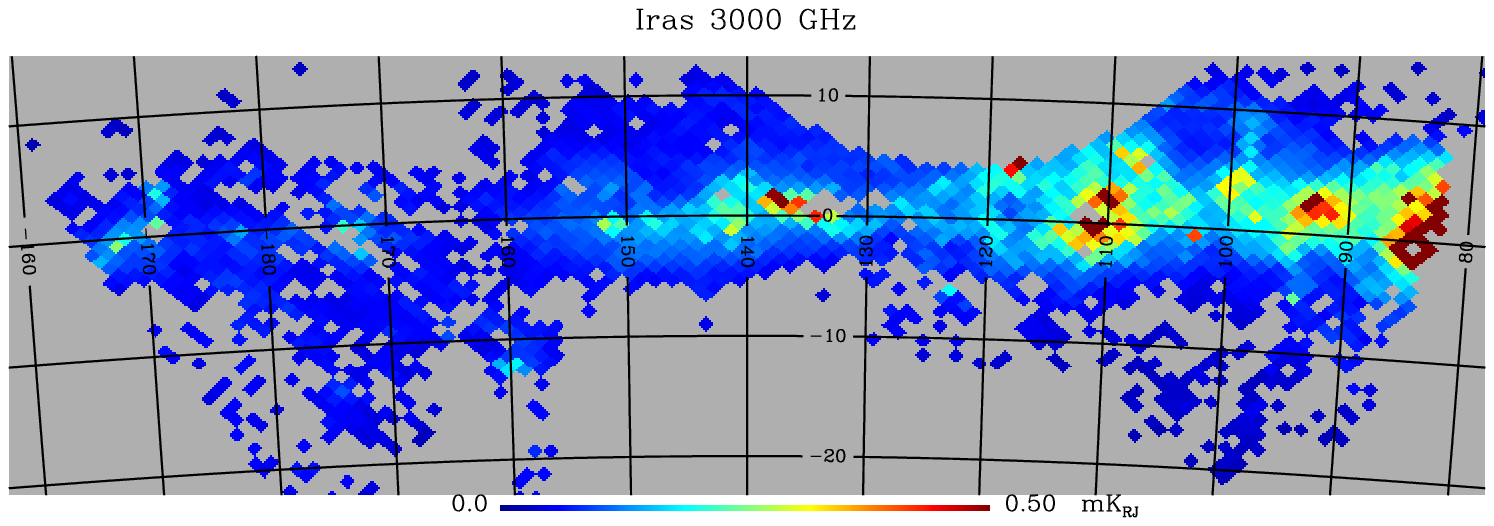}\includegraphics[angle=90,height=4cm,width=6cm]{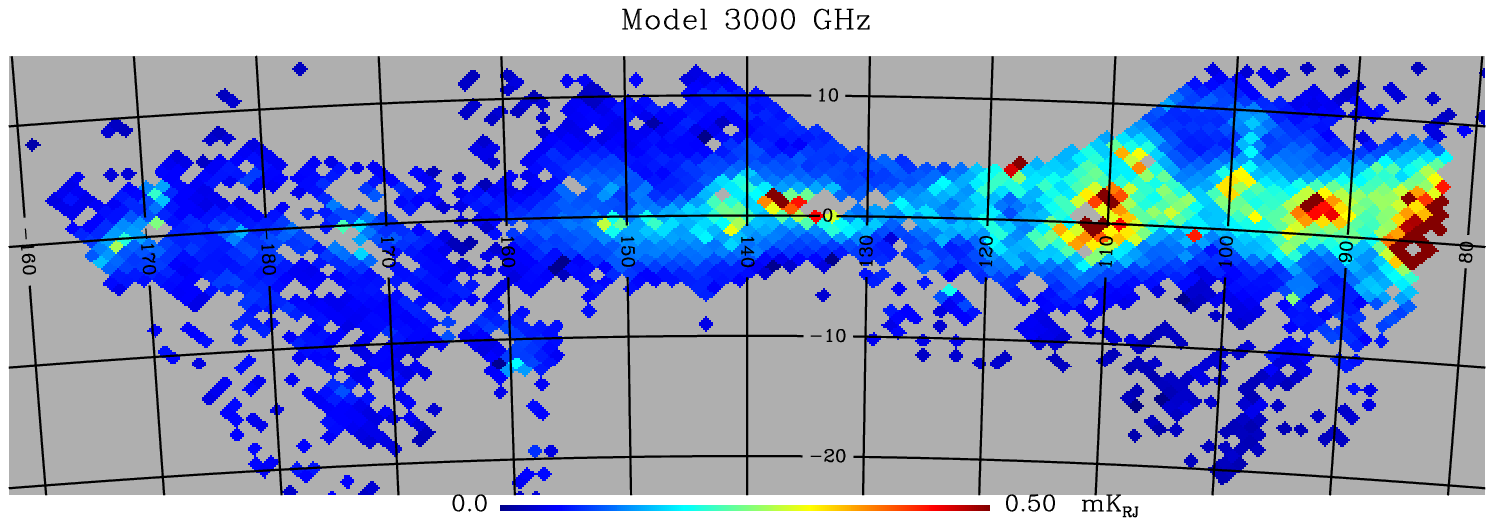}\includegraphics[angle=90,height=4cm,width=6cm]{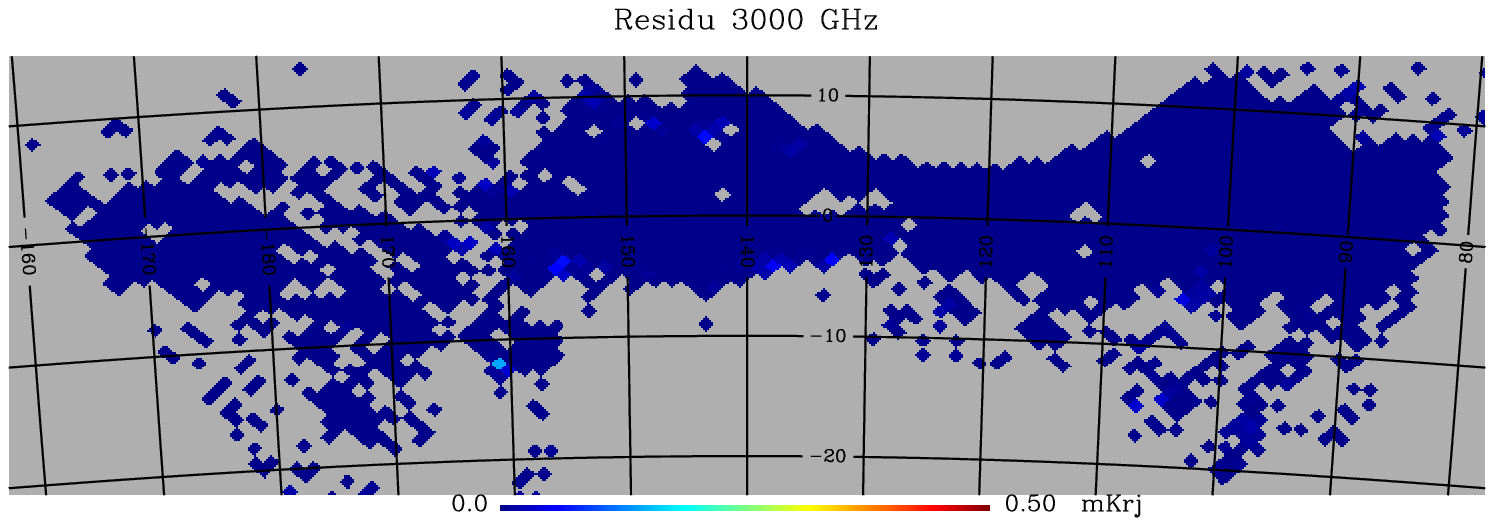}

\includegraphics[angle=90,height=4cm,width=6cm]{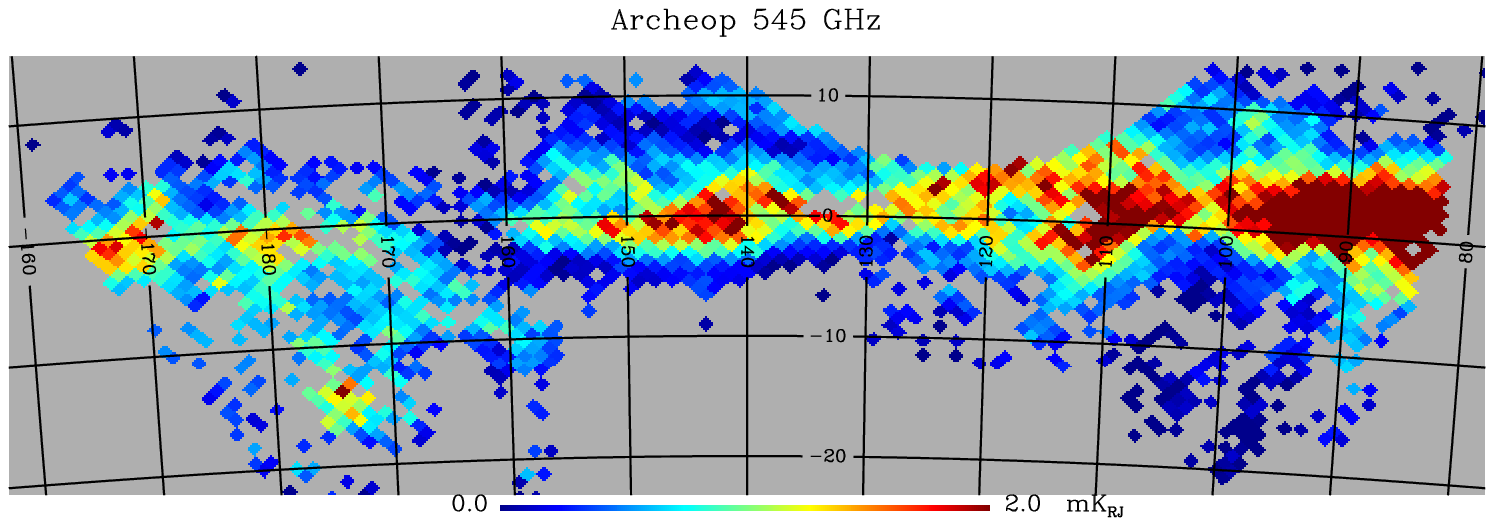}\includegraphics[angle=90,height=4cm,width=6cm]{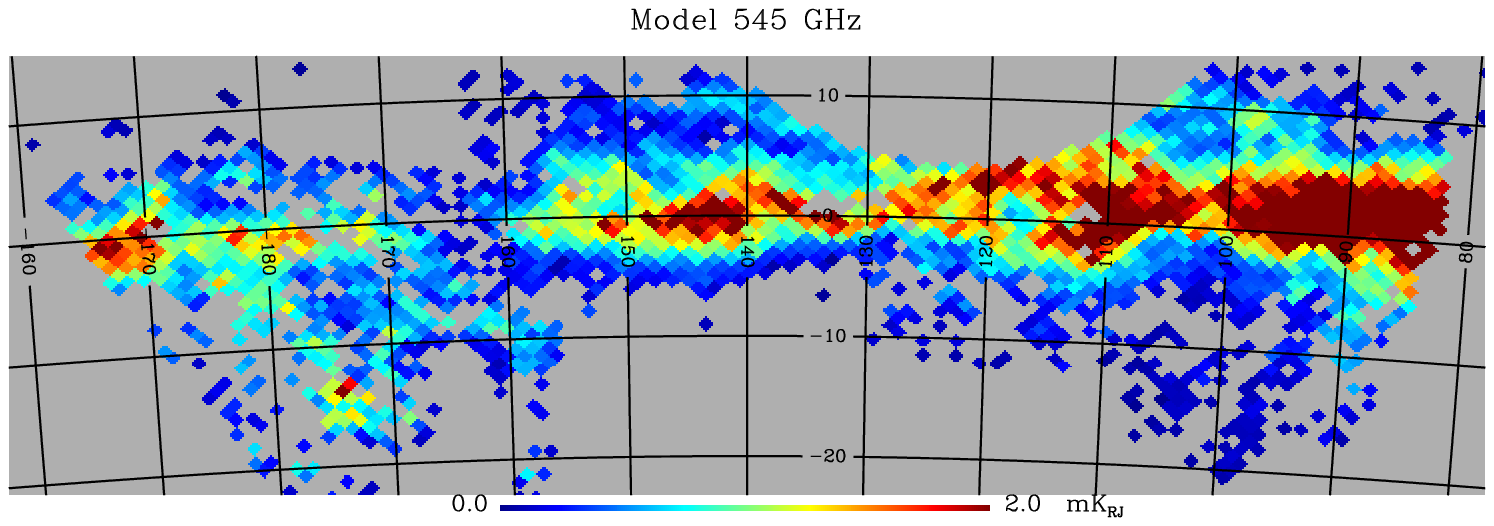}\includegraphics[angle=90,height=4cm,width=6cm]{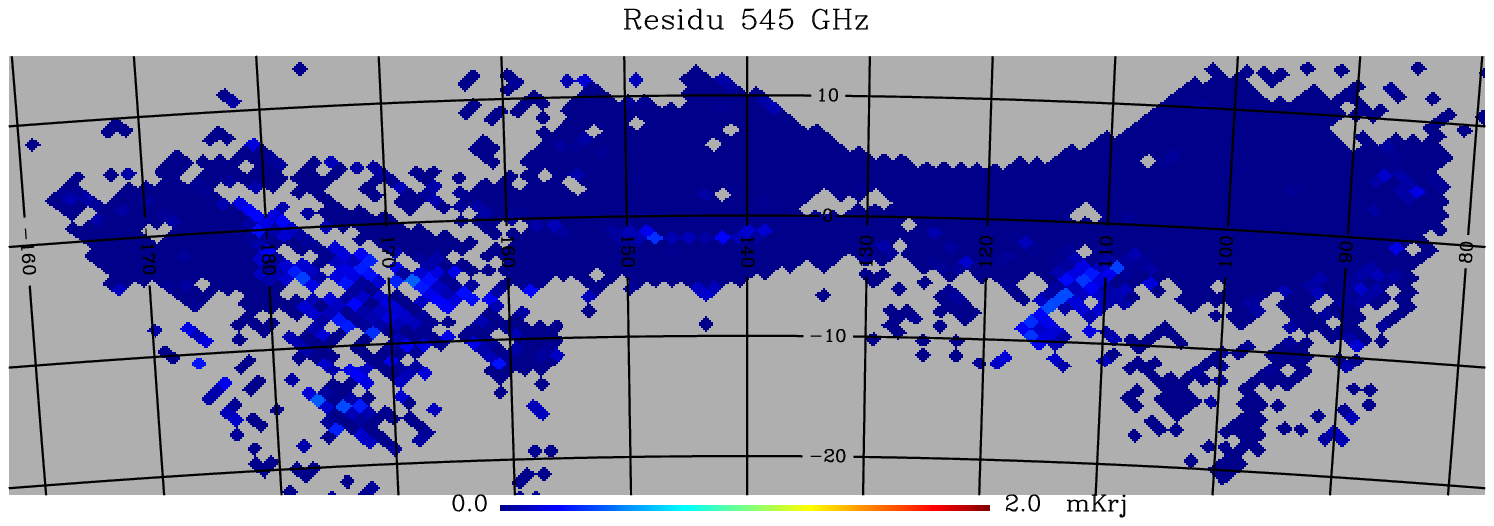}
\includegraphics[angle=90,height=4cm,width=6cm]{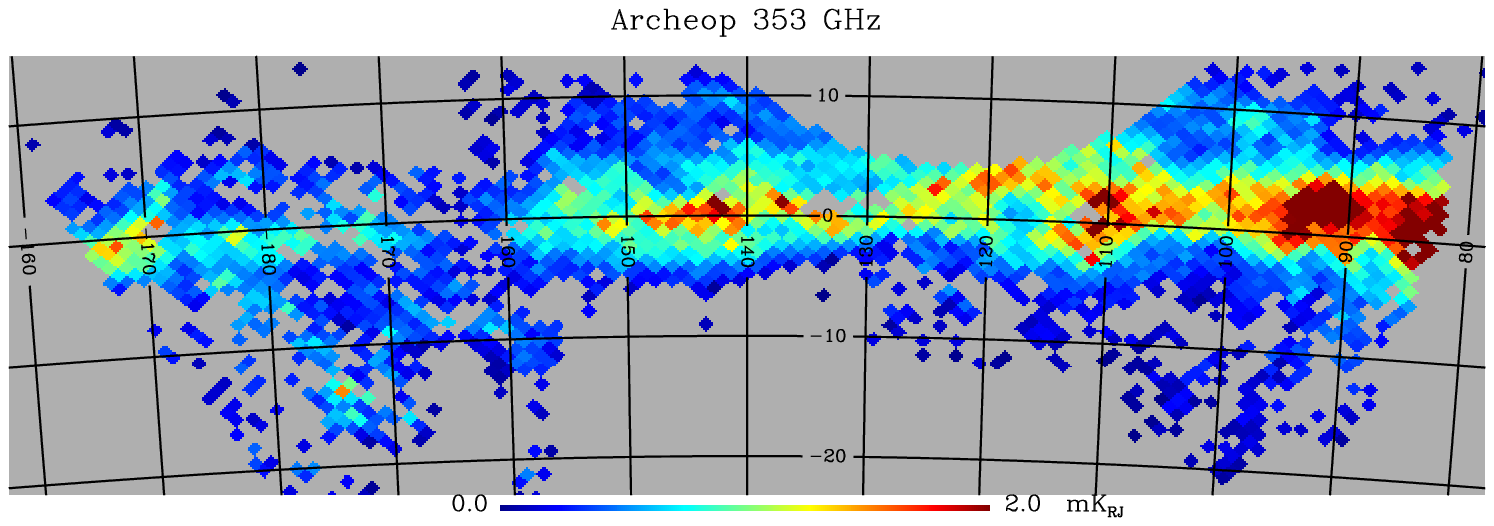}\includegraphics[angle=90,height=4cm,width=6cm]{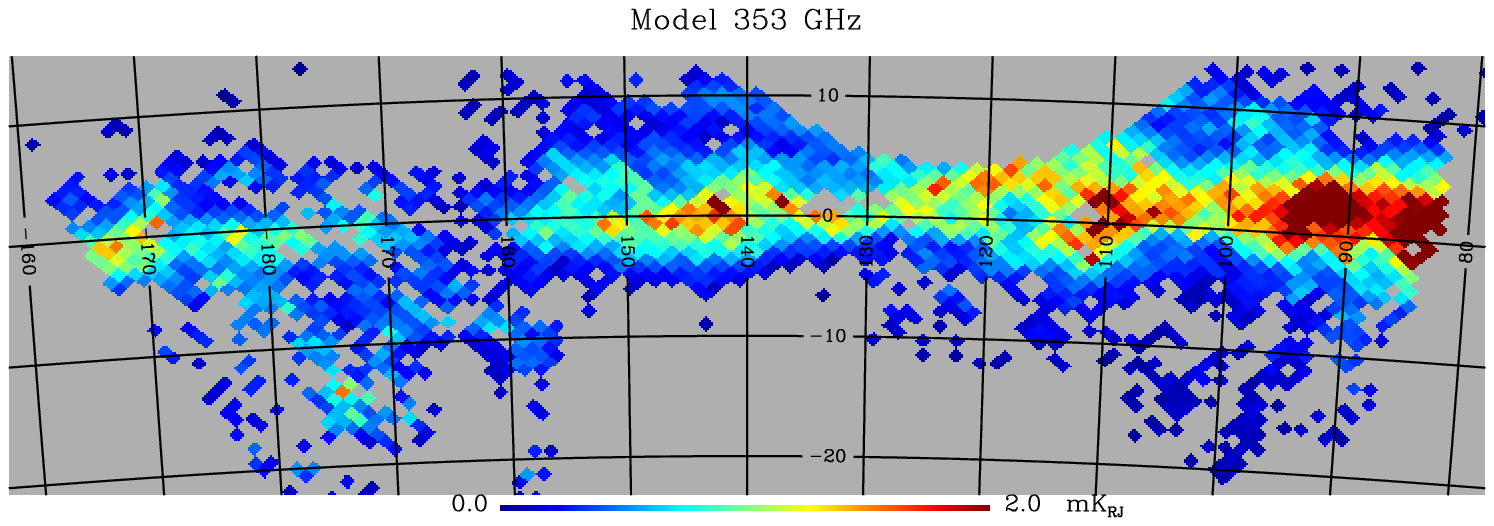}\includegraphics[angle=90,height=4cm,width=6cm]{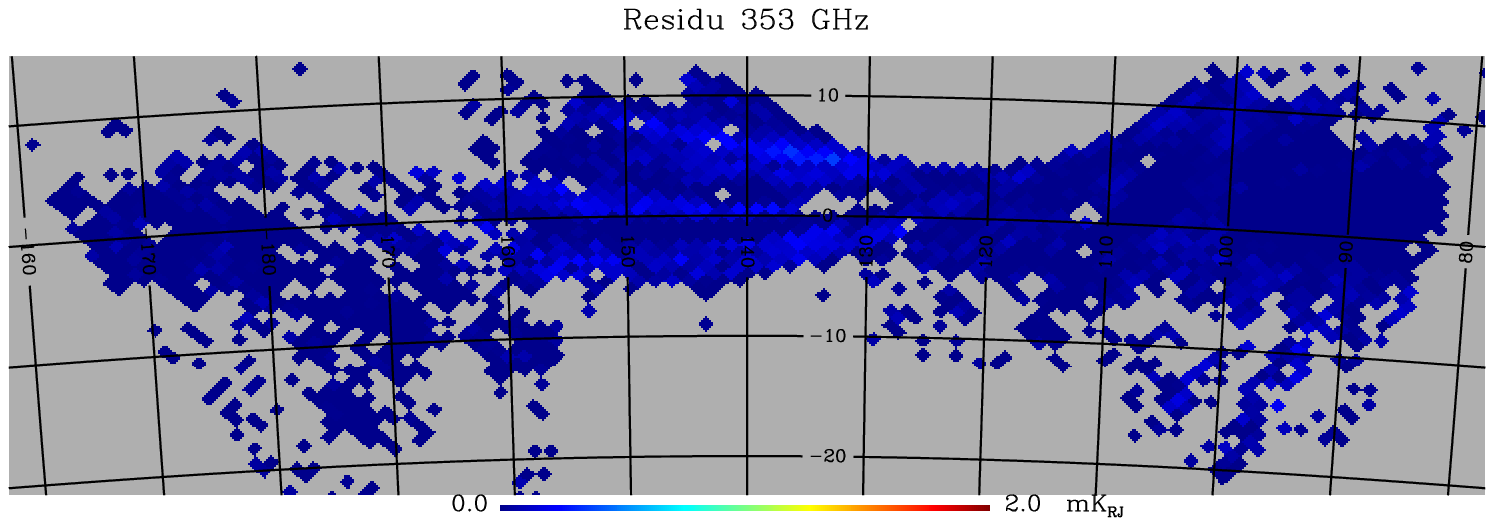}
\includegraphics[angle=90,height=4cm,width=6cm]{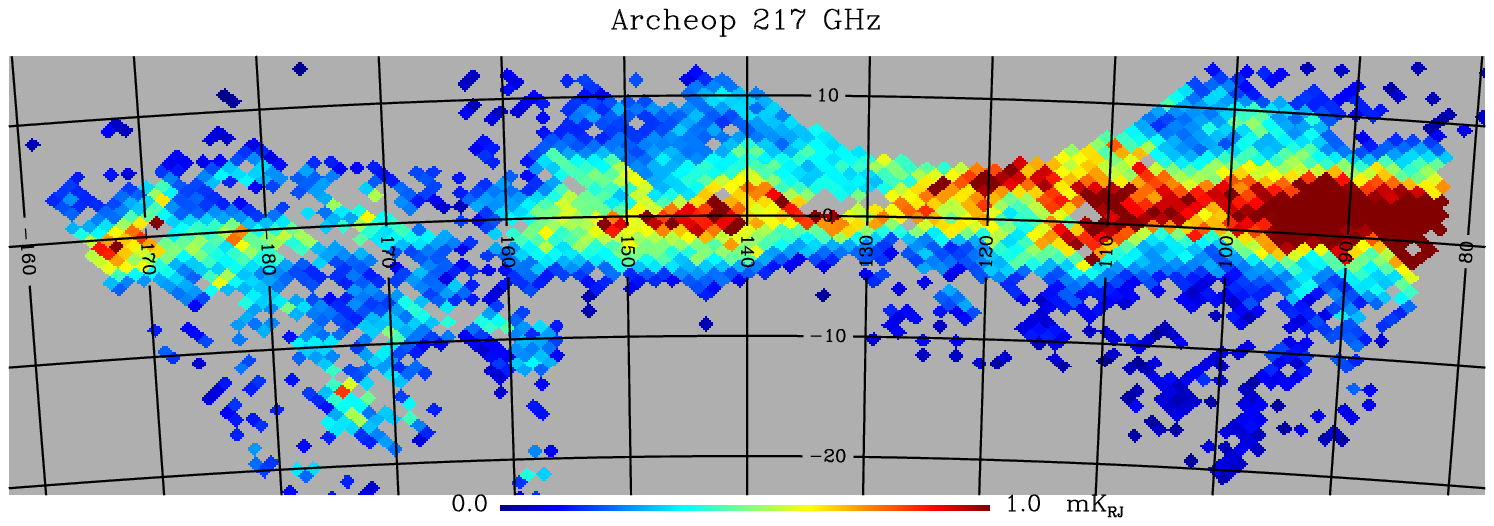}\includegraphics[angle=90,height=4cm,width=6cm]{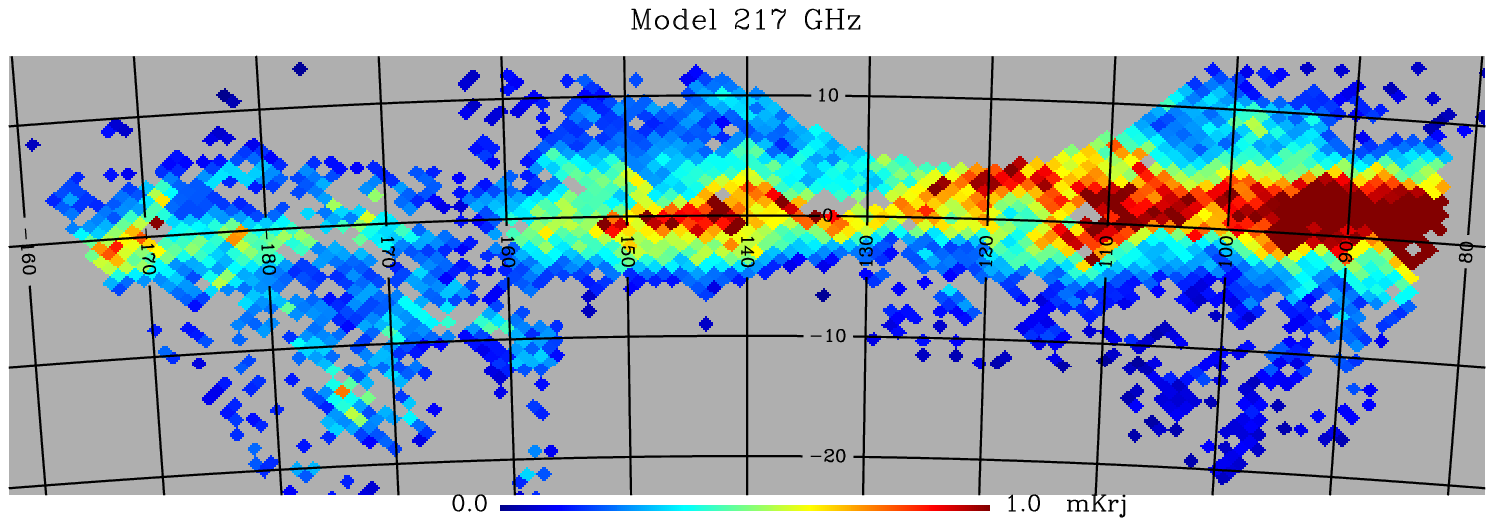}\includegraphics[angle=90,height=4cm,width=6cm]{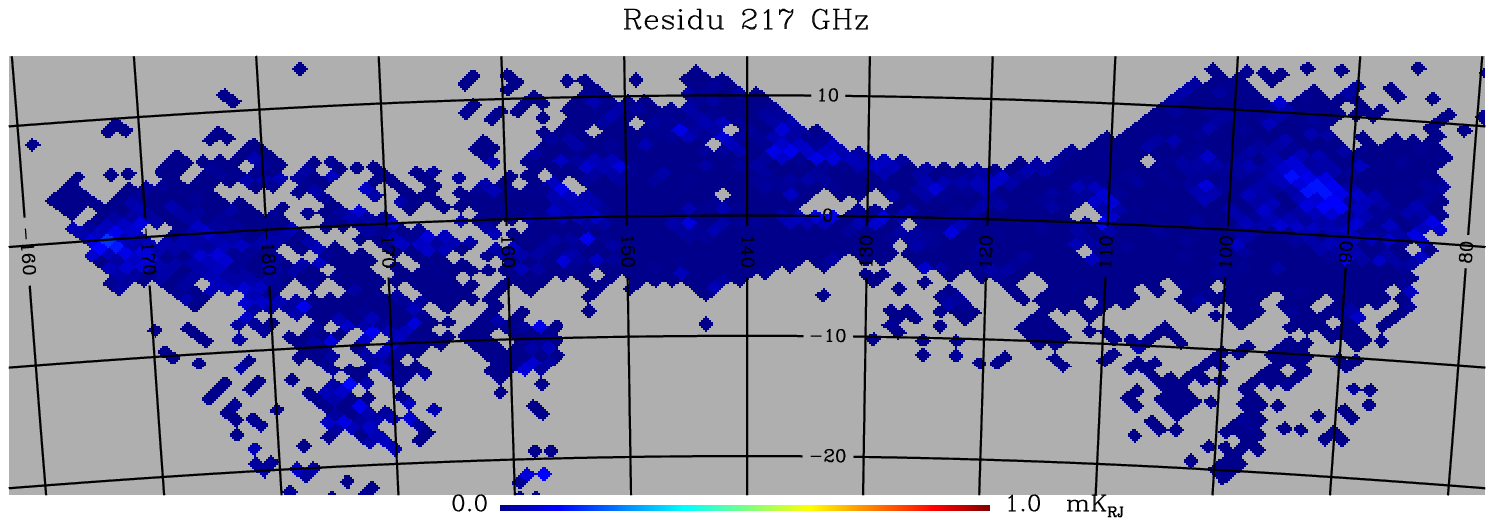}
\includegraphics[angle=90,height=4cm,width=6cm]{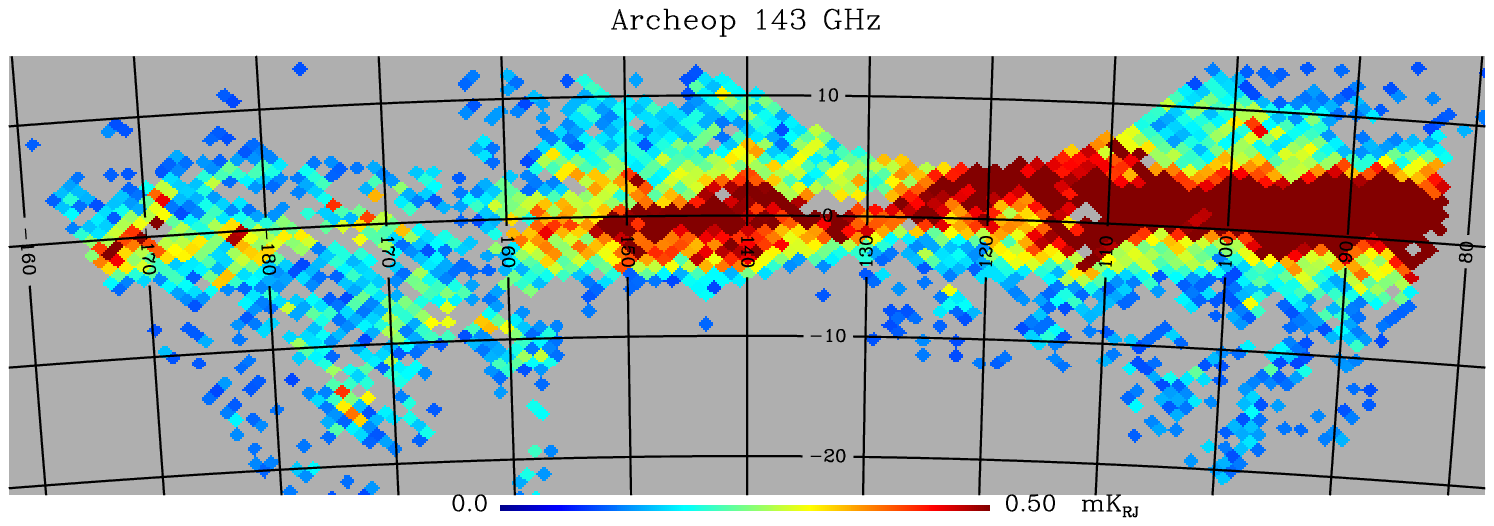}\includegraphics[angle=90,height=4cm,width=6cm]{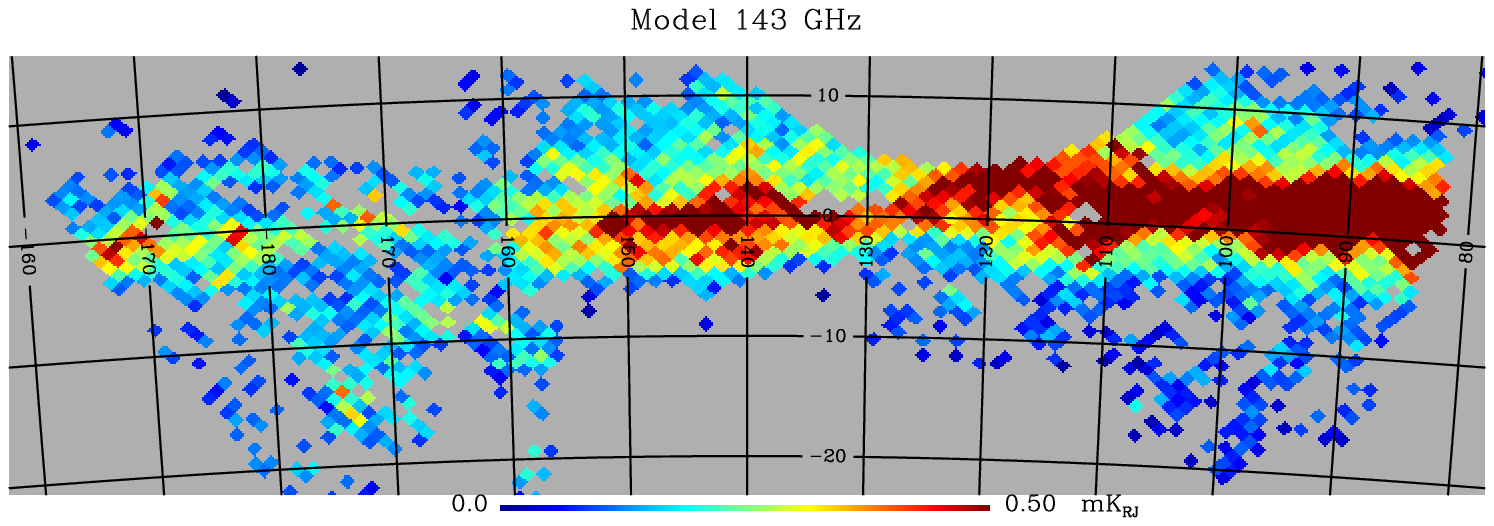}\includegraphics[angle=90,height=4cm,width=6cm]{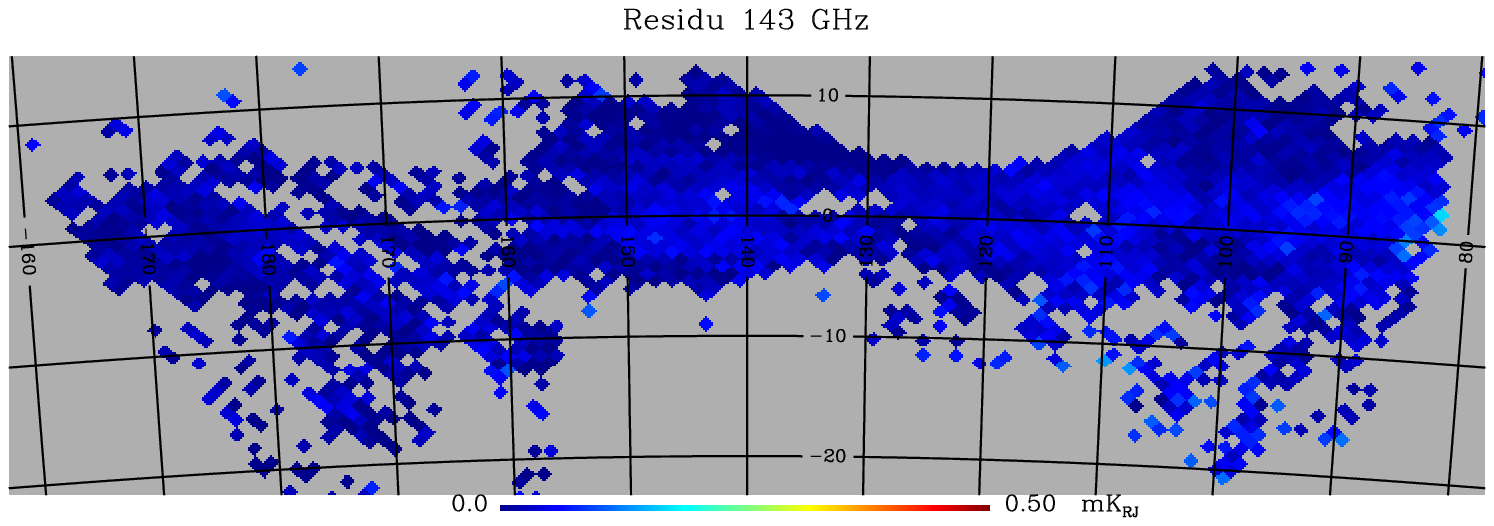}

\caption{Temperature maps (mK$_{RJ}$) for the \Archeops\ data (\emph{left}), the thermal dust emission model \emph{(center)} and residuals \emph{(right)}. From top to bottom
we present the 545, 353, 217 and 143 GHz maps. \label{fig:chp4_dust_nuarch}}
\end{figure*}

We describe in this section the data used for the analysis presented in this paper.
As we are interested in the Galactic diffuse emission we consider only large coverage sky surveys
in the radio, microwave, millimeter and infrared domain including the 408 MHz all-sky survey and
the \Wmap\, \Archeops\ and \Iras\ data.

\subsection*{408 MHz all-sky survey}
In the radio domain, the 408MHz all-sky continuum survey (\cite{haslam}) at a resolution of  of 0.85 degrees, is a good tracer of the  synchrotron emission.
In particular, we use the 408MHz all-sky map available on the LAMBDA website in the \healpix\ pixelisation scheme~\citep{gorski}.
The 408 MHz all-sky survey map was smoothed down to a resolution of 1 degree and downgraded to $N_{side}= 64$ in the \healpix\ pixelisation scheme (\cite{gorski}). 
The uncertainties on this map are assumed to be of 10 \% following~\cite{haslam} and are mainly due to calibration errors.

\subsection*{\Wmap\ }

\indent In order to estimate the diffuse Galactic emission at microwave frequencies we used the maps in temperature using the K, Ka, Q, V and W band maps of the \Wmap\  mission of its 7-years \Wmap\\citep{hinshaw09}. In particular, we used the co-added maps available on the the Lambda web site, also smoothed down to a resolution of 1 degree and downgraded to  $N_{side} = 64$. Uncertainties in the \Wmap\ data were computed assuming a uncorrelated anisotropic noise as described in \citep{hinshaw09}. The variance per pixel at the working resolution was computed using the variance of a single hit and the number of hits per pixels.

\subsection*{\Archeops\ }

\indent In the millimeter wavelengths we use the  \Archeops\  balloon experiment~\citep{benoit2004a} data. \Archeops\ observed the sky at four frequency bands: 143, 217, 353 and 545 GHz, with a resolution of 11, 13, 12 and 18 arcmin respectively\citep{macias2006}. The \Archeops\ survey covers about 30 \% of the sky mainly centered in the Galactic anti-center region.
We use here the original \Archeops\ maps which were also smoothed down to a resolution of one degree and downgraded to  $N_{side} = 64$.

\subsection*{\Iras\ }

\indent In the infrared, we have used the new generation of the \Iras\ data (\emph{InfraRed Astronomical Satellite})
at 100 and 60 $\mu$m (3000 and 5000 GHz). This release of the \Iras\ data is called \Iris\
(\emph{Improved Reprocessing of the \Iras\ data})\citep{miville} and has been built with a better destroying, a better subtraction of the zodiacal light and a calibration and a zero level compatible with the far infrared instrument, FIRAS, of COBE. The IRIS maps were also smoothed down to a resolution of one degree and downgraded to 
$N_{side} = 64$.

\indent In order to avoid the contamination from the CMB at intermediate frequencies, 30-200 GHz, we have 
restricted our study to the Galactic plane where the Galactic emissions dominate over 
the cosmological CMB emission. In practice, we selected those regions in the Archeops 353 GHz map with intensity above
3000 $\mu K_{RJ}$ or higher.  This corresponds to 1391 pixels at $N_{side} = 64$ in the anticenter region.

\begin{figure*}[!h]
\centering
\includegraphics[angle=90,height=4cm,width=5.5cm]{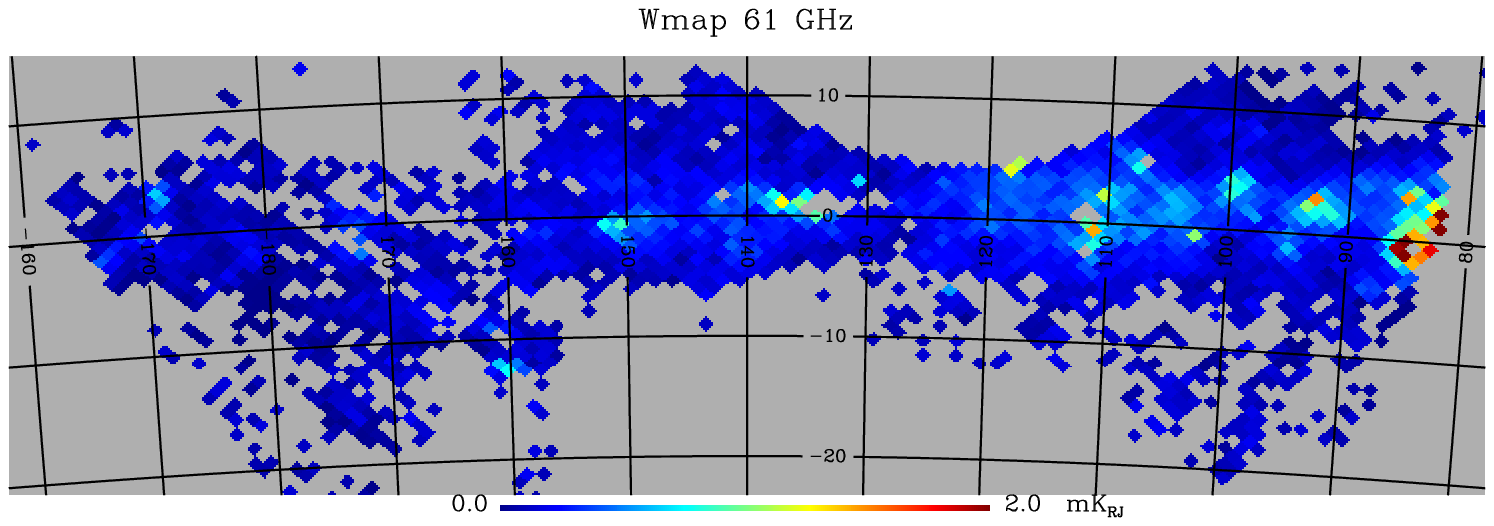} \includegraphics[angle=90,height=4cm,width=5.5cm]{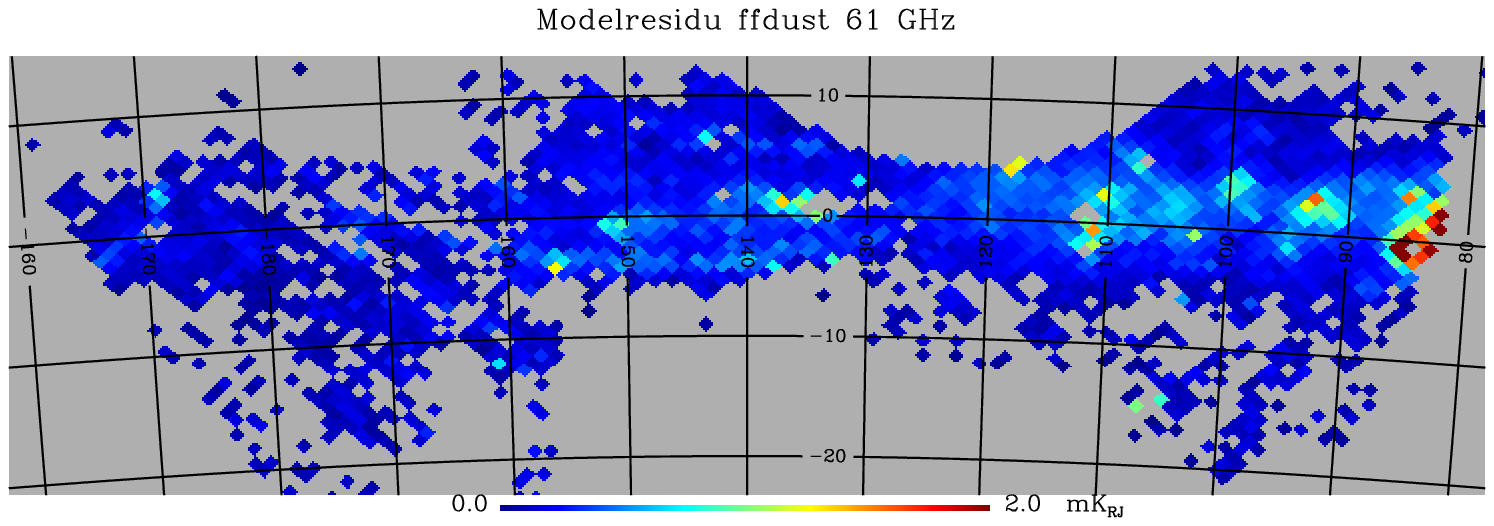} \includegraphics[angle=90,height=4cm,width=5.5cm]{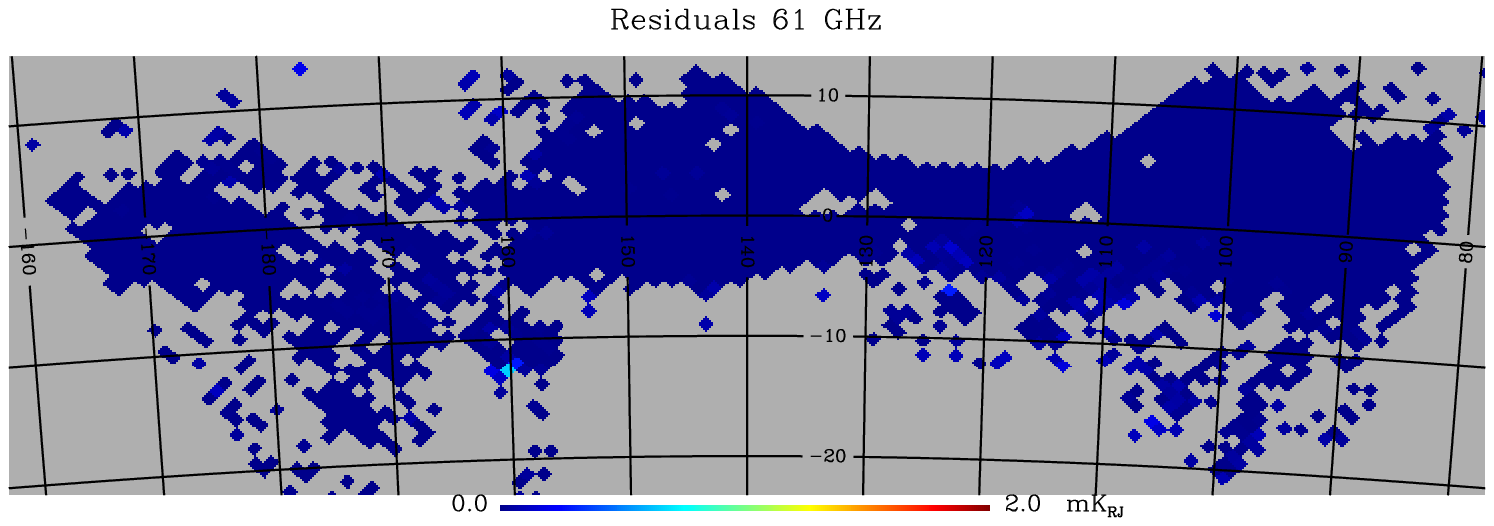}

\includegraphics[angle=90,height=4cm,width=5.5cm]{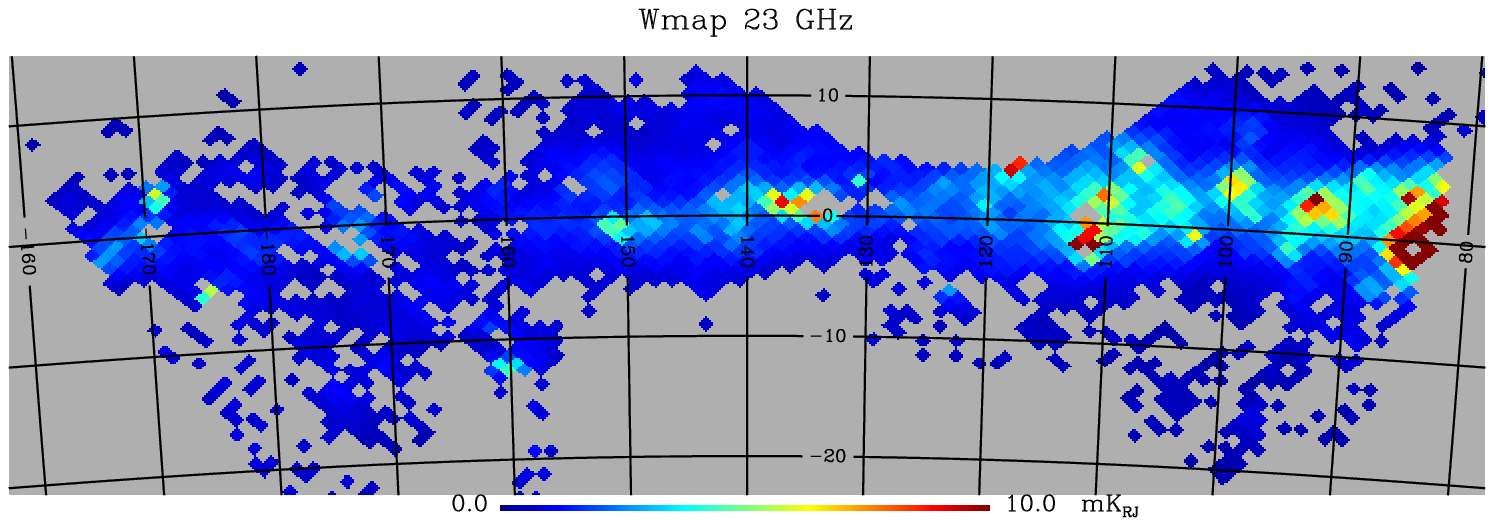} \includegraphics[angle=90,height=4cm,width=5.5cm]{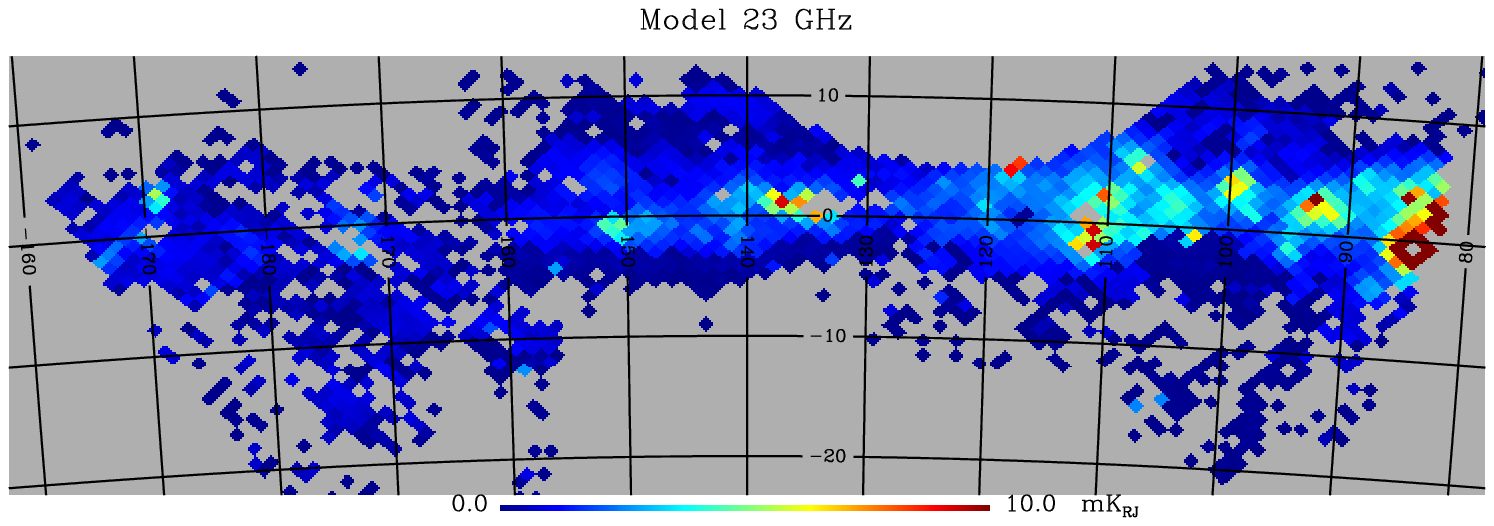} \includegraphics[angle=90,height=4cm,width=5.5cm]{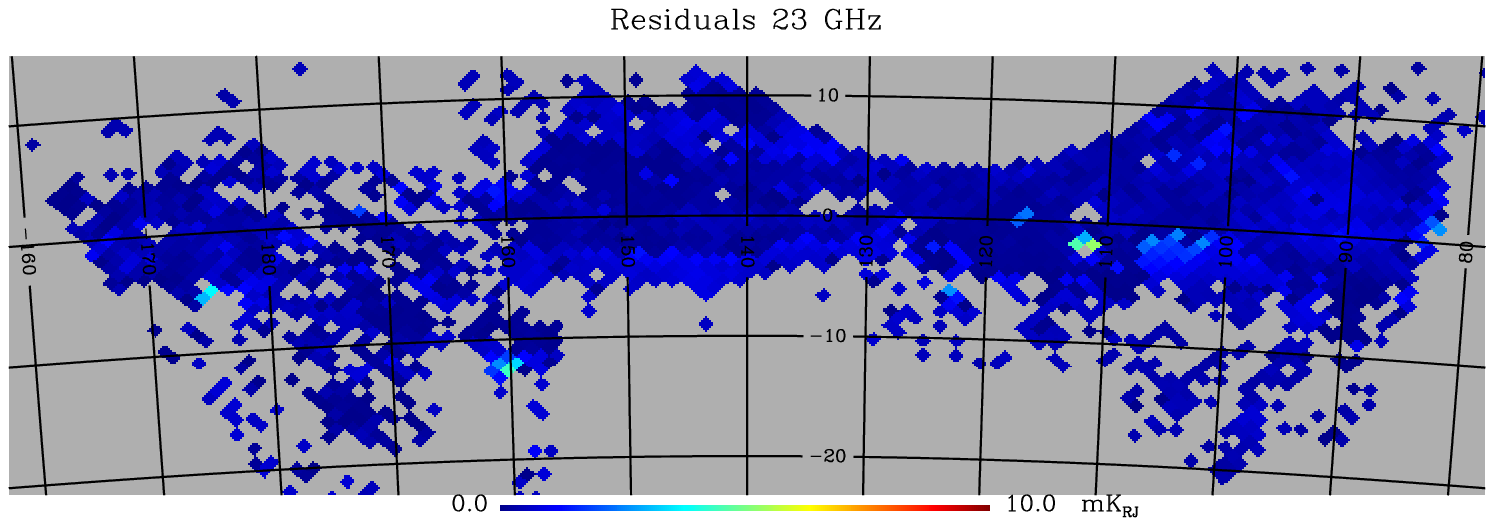}

\caption{From left to right: Residuals after subtraction of the thermal dust, free-free and
anomalous emission models for the 61 (top)  and 23~GHz (bottom) maps.  \label{fig:residu_dust_ff_23Ghz}}
\end{figure*}

\section{Diffuse galactic thermal dust emission}
\label{dust}

\indent We first study the electromagnetic and spatial properties of the thermal dust diffuse Galactic emission.
In order to model the intensity of the thermal dust emission, we use a simple grey body spectrum of the form
\be
\centering
I_{\nu} = I_{0} \nu^{\beta_d}B_{\nu}(T_d)
\ee

\noindent  where $\beta_d$ is the spectral index of the thermal dust emission and $T_d$ is the dust temperature. 
\\


\begin{table}
\begin{center}
\caption{\footnotesize Range of values considered for 
the parameters of the thermal dust emissivity model.\label{tab:4_param_emdust}}
\vspace{0.3cm}

\begin{tabular}{|c|c|c|} \hline
 Parameters  &   Range   &   Step  \\\hline
$\beta_d$ &    $[-1.0, 4.0]$  &   $0.02$   \\\hline
$T_d$  &   $[10.0,37.0]$  &   $0.1$ \\\hline 
\end{tabular}
\end{center}
\end{table}

\indent We used the \Archeops\ and \Iris\  100 $\mu$m maps to characterize the 
dust thermal emission model. We fitted the data to the model pixel by pixel using as free parameters $I_{0}$, $\beta_d$ and
$T_{d}$, and the following likelihood function
\be
- \log \mathcal{L}_{\nu} =  \sum_{\nu}\frac{(D_{\nu}^{p}-M^{p}_{\nu})^2}{{\sigma^{p}_{\nu}}^2}
\ee

\noindent where $D_{\nu}^{p}$ and $M_{\nu}^{p}$ correspond to the data and model at the pixel $p$ within the mask and for the
observation frequency $\nu$ (= 143, 217, 353, 545 and 3000 GHz).  $\sigma_{\nu}^{p}$
is the 1-$\sigma$ error bar associated to $D_{\nu}^{p}$. $\beta_d$ and $T_d$ were explored using an uniformly spaced grid as defined in Table~\ref{tab:4_param_emdust})
while $I_{0}$ was computed via a linear fit for each pair ($\beta_d$ ,$T_d$).
The instrumental noise in the \Archeops\ maps has been estimated using simulations of the noise in the \Archeops\ Time Ordered Information (TOI) following the method
described in \cite{macias2006}. The variance per pixel was calculated from 250 simulated noise maps at one degree resolution
and $N_{side}=64$. The error bars for the \Iris\ data at 100 $\mu$m were set to 13.5 \% following~\citep{miville} as they are dominated by calibration errors. \\ \\
 

\indent 
In Figure~\ref{fig:chp4_Tdust_map} we present maps of the dust temperature and spectral index within the considered
mask. We also show the statistical uncertainties on these parameters. As expected the errors increase significantly on the edges of the maps. These noisy pixels
will be excluded from the analysis hereafter. We can also notice that in the inner regions the statistical errors are significantly smaller than the observed
dispersion for the two parameters. We observe that the mean dust temperature is 20.0 K with 2.1 K dispersion, while the mean
instrumental uncertainties are of the order of 1~K. In the same way, the mean dust spectral index is 1.40 with a dispersion
of 0.25, and the mean instrumental uncertainties of the order of 0.1.  \\


Figures~\ref{fig:chp4_dust_nuiris} and \ref{fig:chp4_dust_nuarch} compare the best-fit thermal dust model to the \Iras\  and \Archeops\ data. From left to right we show
the data, the model and residuals for all frequencies. For the \Archeops\ data the residuals are at most 10 \% of the total intensity.  In the case of the \Iras\ data the model reproduce rather well the 100 $\mu$m map. However,  at 60 $\mu$m, the residuals are important and the model is not able reproduce the structure in the data. Residuals can be as important as 60 \% 
of the total intensity. This can be explained by the presence of a hotter dust component as discussed in~\cite{desert90}. This component is out of the scope of this study
and does not have any consequence in the following study. Table~\ref{tab:dustmodelresiduals} presents the rms of the \Archeops\ and \Iras\ data as well as the  rms of the residuals after subtraction of the dust model. The last column of the table represents the mean standard deviation of the noise in the original maps. We observe that except for the 5000~GHz data the residuals are of the order of magnitude of the noise.

\begin{figure}[!h]
\hspace*{-1.5cm}
\centering
\includegraphics[angle=90,height=6cm,width=8cm]{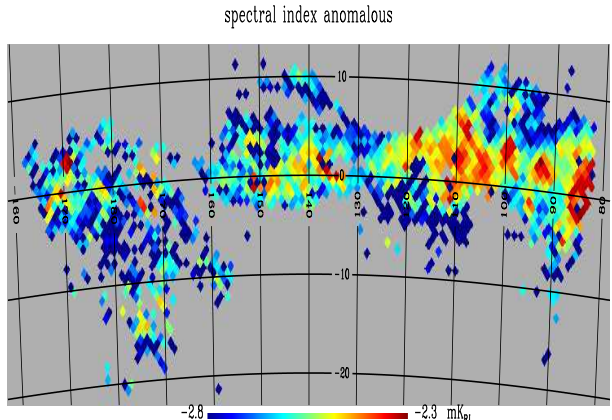}
\caption{Map of the spectral index of the anomalous emission. \emph{(right)}. \label{fig:chp4_betas}}
\end{figure}

\section{Diffuse galactic free-free and synchrotron emissions}
\label{4_ff_model}

\begin{table}
\begin{center}
\caption{\footnotesize Spectral index of the free-free emission at the \Wmap\ frequencies assuming an electronic temperature of $T_e$=8000 K.\label{tab:beta_ff_wmap}}
\vspace{0.3cm}
\begin{tabular}{|c|c|c|c|c|c|} \hline
 central frequency (GHz)  &   23 & 33 & 41 & 61 & 94   \\\hline
 $\beta_{ff}$                &  -2.090  & -2.093 & -2.095 & -2.099  & -2.103   \\\hline
\end{tabular}
\end{center}
\end{table}

\begin{table}
\begin{center}
\caption{\footnotesize Range of the parameters considered for
 the anomalous and free-free emission models.\label{tab:param_ff_sync}}
\vspace{0.3cm}

\begin{tabular}{|c|c|c|} \hline
 Parameters  &   Ranges   &   Step  \\\hline
$\beta_s$ & $[-3.7,-2.3]$  & $0.01$ \\\hline
$T_e (K)$ & $[4000.0, 14000]$  & $1000$ \\\hline
\end{tabular}
\end{center}
\end{table}


\indent In order to estimate the contribution of the diffuse galactic free-free
emission, which is expected to be important at the \Wmap\ bands, 
we use the \emph{Extinction-Corrected Halpha Foreground Template} (H$\alpha$) map built by~\cite{finkbeiner2003}. This map was computed using data from the
\emph{Virginia Tech Spectral line Survey} (VTSS), for the North  and of data from the \emph{Southern H-Alpha Sky Survey Atlas} (SHASSA) for the South sky. Correction factors are applied to take into account dust absorption~\citep{finkbeiner2003}. We started from a map at resolution $N_{side}=512$ and downgraded it, as the other maps, at a resolution of $N_{side}=64$. In order to obtain a template of the free-free emission at 23 GHz using the H$\alpha$
 map, we follow~\cite{bennett}. In antenna temperature units and defining the emission measure as $EM=\int n_e^2 d\mathrm{{\bf n}}$, one can write
\be
\centering
\label{eq:Tff_EM}
T^{ff}_A(\mu K) = 1.44 EM_{cm^{-6}.pc}\cdot \frac{[1+0.22ln(T_e/8000K)-0.14ln(\nu/41\mathrm{GHz})]}{(\nu/41\mathrm{GHz})^2(T_e/8000K)^{1/2}}
\ee

\noindent The intensity of the H$\alpha$ $I(R)$ emission  (in Rayleigh units) is defined by 

\be
\label{eq:IR_EM}
\centering
I(R) = 0.44 EM_{cm^{-6}.pc} \left(\frac{T_e}{8000 K}\right)^{-1/2}
\times \left(1 - 0.34ln\left(\frac{T_e}{8000 K}\right) \right)
\ee
\noindent Thus the intensity of the free-free emission ( in mK$_{RJ}$) is given as a function of the 
intensity of the H-$\alpha$ emission (in Rayleighs) by

\be
\centering
\label{eq:ff_IR}
T_{ff} =
\frac{1.44}{0.44}I(R)\frac{\left(1+0.22ln\left(\frac{T_e}{8000K}\right)
  -0.14ln\left(\frac{\nu}{41\mathrm{GHz}}\right)\right)}{\left(\frac{\nu}{41\mathrm{GHz}}\right)^2(1-0.34ln\left(\frac{T_e}{8000K}\right))}
\ee

\noindent We have extrapolated this free-free emission template at each
of the \Wmap\ frequencies assuming that the electromagnetic spectrum of the
free-free emission is well represented by a power law of the form
$\nu^{\beta_{ff}}$ \citep{bennett}

\be
\centering
\beta_{ff} = -2 - \frac{1}{10.48 + 1.5\ln (T_e/8000K)-\ln(\frac{\nu}{41GHz})}
\ee

\noindent We set a standard value for the electronic temperature at 8000 K, following~\citep{otte}. The values of the spectral index obtained at the \Wmap\ frequencies assuming these hypothesis are given in Table~\ref{tab:beta_ff_wmap}.\\


\begin{table*}
\begin{center}
\caption{rms of the WMAP data and of the residuals after subtraction of the dust, free-free and standard synchrotron
model and of the dust, free-free and anomalous emission model compared to the standard deviation of the noise.\label{tab:dustsyncfreeanomalous}}
\vspace{0.3cm}

\begin{tabular}{|c|c|c|c|c|} \hline
Frequency (GHz) &  Data rms (mK$_{RJ}$)&	Residual DFS rms (mK$_{RJ}$) & Residual DFA rms  (mK$_{RJ}$) & Noise standard deviation (mK$_{RJ}$) \\
\hline
23   &   2.03831   &   2.02387    & 0.353262     & 0.183557 \\
33   &  0.907433   &  0.884012  &  0.0881647  &  0.0612279  \\
41   &  0.562764   &  0.531677  &  0.0372230  &  0.0438113  \\
61   &  0.256907   &  0.205086  &  0.0377750  &  0.0205909  \\
94   & 0.209367    & 0.112932   & 0.0530679  &  0.0239733  \\
\hline	
\end{tabular}
\end{center}
\end{table*}

\indent In order to model the synchrotron contribution we used the
\emph{408 MHz all-sky continuum survey} as a template map. We
extrapolated it at all the considered frequencies assuming a
power law like electromagnetic spectrum in antenna temperature
with  fix spectral index that we set to -2.7~\citep{bennett}.  \\ \\



\indent  In the second column of Table~\ref{ab:dustsyncfreeanomalous} we present the rms of the residuals after subtraction of the Galactic thermal dust, synchrotron and free-free
emission models. These residuals are significant: up to 90 \% of the original emission (first column of the table). We have observed both point like and diffuse
structures in these residuals. The former are more probably related to uncertainties in the modeling of the free-free emission.
By contrast, the extra diffuse emission is most probably related to anomalous emission. This hypothesis is considered
in the following section.

\section{Study of the anomalous emission}
\label{sync}

\indent In the previous section we concluded that the observed emission in the range from 23 to 94 GHz can not
be explained only by the combination of the canonical Galactic diffuse emission: thermal dust, soft synchrotron and
free-free. Indeed, we have observed that in some compact regions there seems to be extra free-free emission
with respect to the predictions from the H$_{\alpha}$ template . Furthermore, the diffuse emission is underestimated in general indicating
either an extra component or a softer synchrotron component. In order to investigate these two problems we have
considered a two component model composed of free-free and anomalous emissions in addition to diffuse thermal dust emission.
We assume that the free-free and the anomalous emissions follow a simple power-law model such that

\be
\centering
M_{\nu} = A_{\mathrm{anom}}\nu^{\beta_\mathrm{anom}} + A_{ff}\nu^{\beta_{ff}(T_e, \nu)} 
\ee 
where $M_{\nu}$ are the observed maps in $K_{RJ}$ units at the frequency $\nu$ after subtraction of the contribution from thermal dust. Finally, we consider 4 free parameters in the model: 
the normalization coefficients $A_{sync}$ and $A_{ff}$, the spectral index $\beta_s$ 
of the anomalous component and the free electron temperature (\ref{eq:eq_Te_ff}). To simplify the fitting procedures we vary $\beta_{s}$ and $T_{e}$ in the ranges shown in Table~\ref{tab:param_ff_sync}.
Notice that we have not explicitly consider the canonical synchrotron 
emission in this model. Indeed, our so called anomalous component will be a mixture of real anomalous emission and canonical synchrotron emission.

\indent
We fit this two-component model to the dust subtracted \Wmap\ maps and to the  408 MHz map for which the thermal dust emission is negligible. 
As discussed before, the uncertainties on the \Wmap\ data have been calculated assuming anisotropic white noise on the maps. We compute the variance r pixel using the variance per single observation provided on the LAMBDA website and maps of the number of hit counts. For the 408 MHz map we assume 10 \% uncertainties as discussed in
Section~\ref{data}. It is important to notice that an alternative three component component model (including free-free, canonical synchrotron and anomalous emission)
would imply at least 6 free parameters to be fitted on only 6 sky maps. That is why we have chosen to consider a two-component model only. \\

\indent 
From the results of the fit we observe that the anomalous emission seems to  dominate the diffuse component at 1 GHz while the free-free emission seems to be mainly located in few compact regions. In Figure~\ref{fig:chp4_betas}  we present the map of the reconstructed spectral index for the anomalous emission, $\beta_{s}$.
 We observe that the anomalous emission seems to be well represented by a power-law with average spectra index of -2.5.
Similar results have been found by \cite{bennett, hinshaw2007} who claim conclusive evidence for hard synchrotron emission. 
In our analysis we did not dispose of data in the frequency range from 10 to 20 GHz to discriminate between this hypothesis and spinning dust emission
(refer to  \cite{draine1998} for a more complete review on spinning dust emission).
It is important to notice that currently spinning dust emission has mainly being found in particular Galactic clouds (see
for example \cite{costa1999, costa2002, lagache2003, hildebrandt2007,watson, ysard, dickinson2010a, bot,2011A&A...536A..20P} ). 
Regarding the electron temperature, we have found that a physically
accessible temperature is associated to only 40 out of 1039 pixels considered. These pixels corresponds to the intense regions on the free-free
map at 1~GHz. For the other pixels, the temperature is higher than the upper limit allowed~\citep{otte} and then can not be linked to the free-free emission.

\section{Summary and conclusions}
\label{conc}

\indent  We have presented in this paper a detailed analysis of the Galactic diffuse emissions at the Galactic anti center in the frequency range from 23 to 545 GHz.
We have shown that a simple grey-body model can be used to describe the thermal dust emission in the frequency range from 100 to 3000 GHz.
We find a mean temperature of 20~K with an intrinsic dispersion of 2.1~K  and a spectral index of 1.4 with intrinsic dispersion of 0.25. 
These values are significantly larger and lower than expected from canonical models of the dust emission, $T_{dust} \sim 17$~K and
$\beta_{dust} = 1.8 - 2$ (see for example \cite{1999ApJ...524..867F,2011A&A...536A..24P}). The same kind of results have been found by \cite{2011A&A...536A..19P}.
although as they fixed the spectral index to 1.8 they obtain a lower temperature of 14~K. We have performed a similar analysis fixing $\beta_{dust}=1.8$
and we have also obtained lower dust temperatures. At high frequencies  (above 3000 GHz) extra hot thermal dust emission from small dust grains is needed to account for the observations~\citep{desert90}.\\

\indent The former dust model have been used to extrapolate the thermal dust emission to microwave frequencies from 23 to 100~GHz. 
After subtraction of the thermal dust emission we have shown that the microwave data can not be simply explained by a combination
of free-free and canonical synchrotron emission.  A more detailed analysis including AME has shown that the latter
can be well approximated by a power-law of average spectral index $-2.5$ in $K_{RJ}$ units. This anomalous emission seems to dominate
the diffuse emission at microwave frequencies while free-free emission seems to be located in few compact regions. Indeed,
we have found that outside those regions the data required electron temperature has not physically meaningful. \\

\indent The spectral index found for the anomalous emission is  consistent with hard synchrotron emission \citep{bennett, hinshaw2007}.
 However, we can not formally conclude on this as our analysis did not include data in the 1 to 20~GHz that would help discriminating this hypothesis
 from spinning dust emission~\cite{draine1998} for which conclusive evidence have been found on some Galactic clouds~\cite{costa1999, costa2002, lagache2003,watson, ysard, dickinson2010a, bot,2011A&A...536A..20P}  and as diffuse emission by~\cite{hildebrandt2007}.

\section*{acknowledgements}
The authors would like to thanks R. Davies and C. Dickinson for useful discussions.
SH would like to thank the LPSC and, especially,  Dr. Juan Macias and Prof. Daniel Santos for the time he spent at LPSC during 2010.

\bibliographystyle{juan} 
\bibliography{biblio} 

\end{document}